\documentstyle[12pt]{article}\textheight 220mm\textwidth 160mm
\begin{document}
\hspace*{11.6cm}KANAZAWA-92-15
\vspace*{1cm}
\begin{center}{\bf Canonical treatment of two
dimensional gravity\\
as an anomalous gauge theory} \end{center}
\vspace*{0.7cm}
\begin{center}
T. Fujiwara$\ ^{(a)}$ and  T. Tabei\\
{\em  Department of Physics, Ibaraki University, Mito 310,
Japan}\\  and\\
Y. Igarashi$\ ^{(b)}$ and K. Maeda\\
{\em Graduate School of Science and Technology,\\ Niigata
University, Niigata 950-21, Japan}\\
and\\
J. Kubo$\ ^{(c)}$\\
{\em  College of Liberal Arts, Kanazawa University, Kanazawa 920,
Japan }
\end{center}
\vspace*{0.7cm}
\begin{center}
ABSTRACT
\end{center}
The extended phase space
method of Batalin, Fradkin and Vilkovisky is applied to
formulate two dimensional
gravity in a general class of gauges.
A BRST formulation of the light-cone gauge is presented to reveal
the relationship between the BRST symmetry and the origin of
$SL(2,R)$ current algebra. From the same principle we derive the
conformal gauge action suggested by David, Distler and Kawai.
\vspace*{1cm}
\footnoterule
\vspace*{4mm}
\noindent
$^{(a)}$ E-mail address: tfjiwa@tansei.cc.u-tokyo.ac.jp  \\
$^{(b)}$ E-mail address: igarashi@ed.niigata-u.ac.jp\\
$^{(c)}$ E-mail address: jik@hep.s.kanazawa-u.ac.jp
\newpage
\pagestyle{plain}

The extended phase space (EPS) method of
Batalin, Fradkin and Vilkovisky (BFV) \cite{bfv,bf} is a powerful
tool in many aspects\footnote{See Ref. \cite{h} for reviews.}.
In particular, it provides us with a general framework for
the gauge-independent investigation on the algebraic
structure of anomalies \cite{fik,fikm}
and also with a canonical treatment \cite{mo,fik1} of
anomalous gauge theories \cite{fs}.

The purpose of this letter
is to show that
Polyakov's string theory \cite{pol2} at non-critical dimensions
\cite{ct} or 2D gravity [11--14] can be consistently treated as an
anomalous gauge theory and that it is possible to formulate the theory
 in most general class of gauges. We thus are able to give a direct
proof that
the light-cone gauge approach of Refs.\ \cite{pol1,kpz} is
gauge equivalent (at
tree level) to the conformal gauge formulation discussed
in  Refs.\ \cite{d,dk}\footnote{There have been various theoretical
indications for the equivalence. See for instance:
Refs.\ \cite{dk,mm}.}.

We shall follow the approach of Refs.\ \cite{mo,fik1}
to anomalous theories \cite{fs}, which
is based on the Hamiltonian formalism
developed by Batalin and Fradkin (BF)
\cite{bf} to quantize systems under second-class constraints.
The main idea of this method \cite{bf} is to
rewrite a system under second-class constraints
in a gauge symmetric way by adding to EPS
compensating fields, the BF fields.
This idea can be simply extended \cite{mo,fik1}
to quantization of anomalous gauge theories \cite{fs},
because they can be understood as second-class-constrained systems.
Remarkably, the re-conversion from the anomalous second-class
constraints back
into the effectively first-class ones can be performed without
considering gauge fixing. Furthermore, we shall use the recent result
\cite{fikm} on the most general form of the BRST anomaly
on the EPS of Polyakov's string theory to carry out the corresponding
 program
for 2D gravity .
We may  therefore expect to arrive at a quantum theory for 2D
gravity formulated in most general class of gauges,
where gauge equivalence is ensured by the
 Batalin-Fradkin-Vilkovisky (BFV) theorem \cite{bfv,h}.

We begin by defining the EPS of Polyakov's theory
described by the classical action\footnote{Strictly speaking, we have to
include a cosmological constant term in (1) to ensure renormalizability.
We have explicitly checked that the presence of the cosmological
constant term
does not alter the essence of our finding, and so we have decided
to suppress that term to avoid inessential discussions in this note.}
\begin{eqnarray}
S_{\rm {X}} &=& -{1 \over 2} \int d^2 \sigma  \sqrt{-g}\,
g^{\alpha \beta}\, {\partial}_{\alpha} X^{\mu}\, \partial_{\beta} X_{\mu}\\
\mbox{with}& &\alpha,\beta=0,1 ~\mbox{and}~ \mu=0,\cdots,D-1,\nonumber
\end{eqnarray}
where we follow the notation of Ref. \cite{gsw}.
The 2D metric variables $
g_{\alpha\beta} $ shall be parametrized by $\lambda^{\pm} \equiv
(\sqrt{-g}\pm g_{01})/g_{11}$, and $\xi \equiv \ln g_{11}$.
At the classical
level, the theory is a system under the five first-class constraints
\begin{equation}
{\pi}^{\lambda}_{\pm}\approx 0 \ ,\  \pi_{\xi}\approx 0~,~
\varphi_{\pm} \equiv \frac{1}{4} (P \pm X^{\prime})^2 \approx 0\ ,
\end{equation}
where ${\pi}^{\lambda}_{\pm}, \pi_{\xi},$ and $P_{\mu}$ denote the
 conjugate momenta
to $\lambda^{\pm}, \xi,$ and $X^{\mu}$, respectively.
 (We use the abbreviations $ \dot{f} = \partial_{\tau} f ~, ~
f^{\prime} = {\partial}_{\sigma}f$, and $\partial_{\pm}
=\partial_{\tau}\pm\partial_{\sigma}$.)
According to these constraints, we define
the EPS by adding
to the classical phase space the ghost-auxiliary field sector
which consists of the canonical pairs
$$
( {\cal C}^{\rm A}~ ,~ \overline{{\cal P}}_{\rm A} )~ ,~
( {\cal P}^{\rm A}~ ,~ \overline{{\cal C}}_
{\rm A} )~ ,~  ( N^{\rm A}\ ,\ B_{\rm A} )~ ,
$$
where  ${\rm A}\  ( = {\lambda}^{\pm},\xi,\pm)$ labels the first-class
constraints in (2).
Then the Becchi-Rouet-Stora-Tyutin (BRST) charge $Q$ can be easily
constructed from the constraints (2) so as to satisfy the super-Poisson
bracket (PB) relation, $\{Q~,~Q\}_{\rm PB} =0$.

In quantum theory, the operator products in $Q$
should be suitably regularized, and
$Q^2$ expressed by the super-commutator $[Q~,~Q]/2$  may fail to
vanish \cite{ko} due to an anomaly.
In Ref.\ \cite{fikm} the most general
form of $Q^2$ in the
EPS has been algebraically derived without specifying
regularization and gauge:
\begin{equation}
Q^2=\frac{i(26-D)}{48\pi} \int d\sigma\, [~{\cal C}^+\,
\partial_{\sigma}^{3}\, {\cal C}^+ -{\cal C}^-\,
\partial_{\sigma}^{3}\, {\cal C}^-~]~.
\end{equation}
This expresses the anomalous conversion
of the nature of the Virasoro constraints $ \varphi_{\pm}$
due to the Schwinger terms.

Given the most general form of the anomaly, we apply
the BF algorithm  to reconvert the anomalous
system back into a gauge symmetric one.
So we introduce a canonical pair of
BF fields $(\theta, \pi_\theta)$ to cancel the anomaly,
and construct new effective Virasoro constraints
 ${\tilde{\varphi}}_{\pm}$ by adding an appropriate polynomial of
the BF fields to $ \varphi_{\pm}$.  This new contribution is
fixed by requiring that the new BRST charge $\tilde{Q}$ involving
${\tilde{\varphi}}_{\pm}$
 becomes nilpotent and reduces to $Q$ when the new fields are set equal to
zero.
One finds that \cite{ct,fik1}
\begin{eqnarray}
\tilde{Q} &=& \int d\sigma[~{\cal C}_{\lambda}^+ {\pi}^{\lambda}_{+}
+{\cal C}_{\lambda}^- {\pi}^{\lambda}_{-}
+{\cal C}^{\xi} \pi_{\xi}\nonumber\\
& &+{\cal C}^+(~{\tilde{\varphi}}_+
+\overline{{\cal P}}_{+}{\cal C}^{+\prime}~)
+{\cal C}^-(~{\tilde{\varphi}}_-
-\overline{{\cal P}}_-{\cal C}^{-\prime}~)\nonumber\\
& &+{\cal P}^{\rm A} B_{\rm A}~  ]
\end{eqnarray}
satisfies the desired requirements, where
\begin{eqnarray}
\tilde{\varphi}_{\pm}&\equiv& \varphi_{\pm} +
\frac{\kappa}{2}\, (~\frac{\Theta_\pm^2}{4}-\Theta_{\pm}^\prime~)\\
\mbox{with}& &
\Theta_{\pm}=\theta^\prime \pm
\frac{2}{\kappa}\pi_{\theta}~\mbox{and}~
\kappa=\frac{(25-D)}{24\pi}~.\nonumber
\end{eqnarray}
We have taken into account the fact that the BF
fields non-trivially contribute to the commutator anomaly.
It requires a renormalization of the anomaly coefficient by $1/24\pi$.

Note that
the BRST charge (4) is obtained prior to gauge fixing.
The gauge fixing appears in defining the total Hamiltonian $~H_{\rm T}$.
Since the canonical Hamiltonian
vanishes in the present case,
 it is given by the BRST variation of the gauge fermion ${\Psi}$,
i.e.\  $H_{\rm T}=-i[\, \tilde{Q}~,~\Psi\, ]$.
The BRST invariant gauge-fixed action of
the theory (1) can be then written as
\begin{eqnarray}
S &=&\int d^2\sigma\, (~{\pi}^{\lambda}_{+}\dot
\lambda^++{\pi}^{\lambda}_{-}
\dot\lambda^-+\pi_\xi
\dot\xi+\pi_\theta\dot\theta+P_{\mu}\dot X^{\mu}\nonumber\\
& & +\overline{\cal P}_{\rm A} \dot{\cal C}^{\rm A}~)
+i\int d\tau\, [\, \tilde{Q}~,~\Psi\, ]~.
\end{eqnarray}
(We have canceled the Legendre term
$\overline{\cal C}_{\rm A}\dot{\cal P}^{\rm A}
+B_{\rm A} \dot{N}^{\rm A}$ in (6) by shifting
the gauge fermion as ${\Psi}\rightarrow{\Psi}+\int d\sigma\,
\overline{\cal
C}_{\rm A} \dot N^{\rm A}$.)
The gauge fermion $\Psi$ is arbitrary so far.
  The BFV theorem \cite{bfv,h} ensures that physical quantities in
the quantum
theory, based on the action (6) along with $\tilde{Q}$, are
$\Psi$-independent.
The nilpotency of $\tilde{Q}$ implies the BRST invariance of
the action (6), and
is essential  for the proof of their theorem \cite{bfv,h}.

It is not straightforward to
recognize 2D gravity in the action (6) because
the geometrical meaning of the 2D metric variables is lost.
To recover it,
we must go to the configuration space.
This requires
elimination of various phase space variables by means
of the equations of motion, and for that we have to specify a gauge.
  In the standard form of the gauge fermion
\begin{equation}
\Psi=\int d\sigma(~\overline{\cal C}_{\rm A}\, \chi^{\rm A}
+\overline{\cal P}_{\rm A}\, N^{\rm A}~)~,
\end{equation}
we have five gauge conditions $\chi^{\rm A}$.
In order to identify the 2D metric variables as well as the
reparametrization
ghosts and the Weyl ghost, we use two of them to impose the
geometrization condition \cite{fikm}
\begin{equation}
{\chi}_{\lambda}^{\pm}=\lambda^{\pm}-N^{\pm}~,
\end{equation}
while making  an (inessential) assumption that
$\chi^{\pm}$ and $\chi^{\xi}$ do not contain
$${\pi}^{\lambda}_{\pm},
\pi_{\xi}, \overline{{\cal C}}^{\lambda}_{\pm},\overline{{\cal P}}_{\pm},
N_{\lambda}^{\pm},N^{\xi},
\overline{{\cal P}}^{\lambda}_{\pm}~,~
\mbox{and}~ B^{\lambda}_{\pm}~.$$
One finds that
\begin{eqnarray}
\lambda^{\pm}&=&N^{\pm}~,
N_{\lambda}^{\pm} = ~ \dot\lambda^{\pm}~,~  N^{\xi} =~ \dot\xi~,
\nonumber\\
{\cal P}^{\pm}&=&{\cal C}_{\lambda}^{\pm} =~ \dot{\cal C}^{\pm}
\pm{\cal C}^{\pm}\partial_{\sigma}N^{\pm}
\mp\partial_{\sigma}{\cal C}^{\pm}N^{\pm} ~,
\end{eqnarray}
can be still unambiguously derived. Then one can verify
that the covariant
variables can be defined as \begin{eqnarray}
C^{0}&\equiv&{\cal C}^{0}/N^{0}~,~C^{1}={\cal C}^{1}-
N^{1}{\cal C}^{0}/N^{0}~,\\
C_W &\equiv&{\cal
C}^\xi-C^{0}N^{\xi}-C^{1}\partial_{\sigma}\xi\nonumber\\
& &-2\,\partial_{\sigma}C^{0}N^{1}-2\,\partial_{\sigma}C~,\\
g_{\alpha\beta}&\equiv&
\left( \begin{array}{cc} -N^{+}N^{-} & (N^{+}
-N^{-})/2 \\
(N^{+}-N^{-})/2 & 1 \end{array} \right)\exp\xi~,
\end{eqnarray}
with the BRST transformation rules \cite{fuji}
\begin{eqnarray}
\delta g_{\alpha\beta}&=&C^{\gamma}\, \partial_{\gamma} g_{\alpha\beta}
+\partial_\alpha C^{\gamma}\, g_{\gamma\beta}\nonumber\\
& &+\partial_\beta C^{\gamma}\, g_{\alpha\gamma}+C_W\,
 g_{\alpha\beta}~,\\
\delta C^{\alpha} &=&C^{\gamma}\, \partial_{\gamma} C^{\alpha}~,~
\delta C_{W} =C^{\gamma}\, \partial_{\gamma} C_{W}~,
\end{eqnarray}
where $ M^{\pm} = M^{0} \pm M^{1}$, and $~\delta f=i[\,\tilde{Q}~,~f\,
]$.  Not that the $\tau $-derivatives in (13) and (14) are
exactly those which appear in
Eqs.\ (9).

At this stage, one is left with three unspecified gauge conditions,
 which correspond to two reparametrization
and one Weyl symmetries.
  We shall consider two gauges below
to illustrate our formulation of 2D gravity,
which would clarify the relations to other approaches.
\newline
(i) Light-cone gauge
\newline
This gauge, $g_{+-} = -1/2,~g_{--} = 0$, is realized by
\begin{eqnarray}
\chi^{+}&=&N^{+}-1~, \quad
\chi^{-} = [(N^{-}+1)e^\xi]~ - 2~,\nonumber\\
\chi^{\xi}&=&\xi-\theta~,
\end{eqnarray}
along with (8).
We substitute the gauge fermion
(7) with $\chi$'s given in (8) and (15) into the action
(6), and then eliminate again all the non-dynamical fields
such as $\pi_\theta,~P_{\mu},~\overline{{\cal P}}_{\pm},~{\cal{C}}^{\xi}
,~\overline{{\cal C}}_{\xi}$ to obtain the gauge-fixed action
\begin{eqnarray}
 S_{\rm LG}  & = &\int d^{2}\sigma\,\{ ~ \frac{1}{2}\,~
                  [(~ g_{11}-1) \, \partial_{-}X^{\mu}\,
\partial_{-}X_{\mu}
                   + \partial_{-}X^{\mu}\, \partial_{+}X_{\mu} ~)
\nonumber\\
 &  &+\frac{\kappa}{4 g_{11}}\ \, ~
                  [~ (~ \partial_{-}g_{11})^2
                   -  2\, (~ \partial_{-}g_{11}~)(~ \ln g_{11}
 ~)^{\prime}
                   +  4\, (~ \ln g_{11}~)^{\prime \prime}  ~]
\nonumber \\
&  &-\overline{{\cal C}}_{+}\, \partial_{-}{\cal C}^{+}
-g_{11}\, \overline{{\cal C}}_{-}\, \partial_{-}{\cal C}^{-}
-2\, \overline{{\cal C}}_{-}\, \partial_{\sigma}{\cal C}^{+}
-\overline{{\cal C}}_{-}\, ({\cal C}^{+} +{\cal C}^{-})\,
\partial_{-}g_{11}~\}~.
\end{eqnarray}
This light-cone gauge action is local, and should be compared
with the non-local action of Polyakov \cite{pol1}.

The equations of motion which follow from this action read:
\begin{eqnarray}
\partial_{-}\partial_{+}X^{\mu}&=&
-\partial_{-}(\, (g_{11}-1)\partial_{-}X^{\mu}\, )~,\nonumber\\
\partial_{-}{\cal C}^{+} &=&0~,\quad
g_{11} \partial_{-}{\cal C}^{-} = -\partial_{+}{\cal C}^{+}
-({\cal C}^{+}+{\cal C}^{-})~\partial_{-}g_{11}~, \nonumber\\
\partial_{-}\overline{{\cal C}}_{-}&=&0~,\quad
\partial_{-}\overline{{\cal C}}_{+}+\partial_{+}\overline{{\cal C}}_{-}=
\overline{{\cal C}}_{-}\, \partial_{-}g_{11}~,
\end{eqnarray}
and also from the variation of $g_{11}$,
\begin{eqnarray}
 \frac{\kappa}{4}~\partial_{-}^{2}\, g_{11}
 & = &   \frac{g_{11}}{4}\, \partial_{-}X^{\mu}\,
\partial_{-}X_{\mu}~+~\frac{\kappa}{2g_{11}}\,
         [~\frac{({\Theta^0_+})^2}{4}~-~({\Theta^0_+})^{~ \prime} ~]\ ,
\end{eqnarray}
where $\Theta^0_+ \equiv \frac{2~g_{11}^{\ \ \prime}}{g_{11}}~-~
\partial_{-}g_{11} $.

The fundamental identity in the light-cone
gauge \cite{pol1,kpz},
\begin{equation}
  \partial_{-}^{3}~g_{++} = 0 ~,
\end{equation}
can be derived as follows.
One first convinces oneself
that in the light-cone gauge the right-hand-side of (18)
is exactly contained in
 $\tilde{\varphi}_{-}/g_{11}$
where
 $\tilde{\varphi}_{-}$
is given in (5).
Then one considers the BRST variation of the first equation of
Eq.\ (17) to obtain
\begin{eqnarray}
 0 & = & ~i ~ \partial_{-}[~ \, \tilde{Q}~,~\overline{{\cal C}}_{-}\, ]
     =   -~ \partial_{-}\, ~  \delta ~
              (\, \overline{{\cal P}}_{-}\,/g_{11} )    \nonumber\\
   & = & \partial_{-}\,~ \{~\frac{\tilde{\varphi}_{-}}{g_{11}} +
\cdots~ \}~,
\end{eqnarray}
where $\cdots$ indicates terms containing ghosts.
Using the equations of motion in (20),
especially those for anti- and ghost fields, one finds that
$$\partial_{-}(\mbox{right-hand-side of (18)})=0 ~.$$
This yields the identity
(19), which is the origin of a $SL(2,R)$ current algebra as shown
by Polyakov \cite{pol1}. So, what we have delivered here
is a BRST formulation of 2D gravity in the light-cone gauge,
thereby clarifying the origin of the $SL(2,R)$ current algebra.
The gauge-fixed BRST charge in this gauge can be explicitly shown
to reduce to
the one given in \cite{kim}.
\newline
(ii) Conformal gauge
\newline
The conformal gauge is defined by (8) and
\begin{equation}
\chi^{\pm}=N^{\pm}-\hat{N}^{\pm}~,
{}~\chi^{\xi}=N^{0} e^{\xi} - {\hat{N}}^{0} e^{\hat{\xi}}~,
\end{equation}
where $\hat{N}^{\pm}$ and $\hat{\xi}$ are background fields which define
a background metric $\hat{g}_{\alpha\beta}$ (see Eq.\ (12)).
We substitute the gauge fermion $\Psi$ (7) with
the gauge-fixing functions (8) and (21) into the action (6) to obtain
the gauge-fixed action.
The momentum variables ${\pi}_{\theta},~P_{\mu}$, and
$\overline{{\cal P}}_{\pm}$
can be eliminated by means of
 the equations of motions, $\pi_\theta = {\pi_\theta}
(\dot{\theta},~\theta,~N^{\pm}),~P_{\mu} =
P_{\mu}(\dot{X},~X,~N^{\pm})$, and
$\overline{{\cal P}}_{\pm} = -\overline{{\cal C}}_{\pm}$.
We next define the covariant anti-ghosts
$\overline b_{\alpha \beta}$ (symmetric and traceless)
and the Weyl anti-ghost
$\overline C_W$ in terms of the BFV anti-ghosts, $\overline{\cal
C}_{\pm}$ and ${\overline{\cal C}}_{\xi}$.
We then arrive at the conformal gauge
action in the configuration space:
\begin{eqnarray}
S_{\rm CG}&=&S_{X}+S_{\phi}+S_{g}\nonumber\\
& &-\int d^{2}\sigma\, \{~B_{\xi}\, (N^{0} e^{\xi} - {\hat{N}}^{0}
e^{\hat{\xi}})
+B_{+}\, (N^{+}-\hat{N}^{+})\nonumber\\
& &+B_{-}\, (N^{-}-\hat{N}^{-})
+ \sqrt{-g}\,[~g^{\alpha\gamma}
\, \overline{b}_{\alpha\beta}\, \nabla_{\gamma}C^{\beta}\nonumber\\
& &+\overline C_{W}\, (C_{W}+\nabla_{\alpha}C^{\alpha})~]~\}\\
\mbox{with}& &
\phi\equiv\theta-\xi~,\nonumber
\end{eqnarray}
where $\nabla_{\alpha}$ is the covariant
derivative, $R$ is the curvature scalar, and
\begin{eqnarray}
S_{\phi} &=&-{\kappa\over2}\, \int d^{2}\sigma\, \sqrt{-g}\,
[~{1\over2}\, g^{\alpha\beta}\,
\partial_\alpha \phi\,
\partial_\beta\phi+R\, \phi~]~,\nonumber\\
S_{g}&=&\frac{\kappa}{2}
\int d^{2}\sigma\sqrt{-g}\, [~
{1\over2}\, g^{\alpha\beta}\,
\partial_\alpha\xi\, \partial_\beta\xi
-R\, \xi\nonumber\\
& &-2\frac{g_{11}}{g}\, \{({g_{01}\over
g_{11}} )^\prime\}^2~] ~.
\end{eqnarray}
The action (22) contains two Liouville-type modes, $\phi$ and
$\xi$.
The BRST transformation of $\phi$ is covariant and given by
\begin{equation}
\delta \phi = C^{\alpha}\partial_\alpha \phi  - C_W,
\end{equation}
and so  it plays the roll of the conformal degree of freedom.
On the contrary to $\phi$, the $\xi\,(=\ln g_{11})$ is a non-covariant
object, and
$S_{g}$ is a  non-covariant expression.
The origin of $S_{g}$ is related to
the fact that the manifest 2D covariance is violated
in the class of regularization schemes we approve.
(The normal-ordering prescription, for example, is such a scheme.)
In order to restore the 2D covariance,
one has to add an appropriately chosen non-covariant counter term
to the action.
The $S_{g}$ is nothing but this counter term action.
Except for the renormalization of the coefficient, it exactly
coincides with the
counter term found in Ref.\ \cite{fikm}, which generates the
coboundary term to covariantize the $Q^2$.
This means that the present approach has
a built-in mechanism to keep the 2D covariance\footnote{In the
operator language,
$S_{X}+ S_{g}$ is thus reparametrization invariant, but not
Weyl invariant. It is the Liouville action $S_{\phi}$ that acts
as a Wess-Zumino-Witten term to recover the Weyl symmetry.}.
We can indeed verify that, if we begin with the
covariantized expression
for $Q^2$ given in Ref.\ \cite{fikm}, we end up with $S_{\rm CG}$
with $S_{g}$ suppressed. This implies
the formal
equivalence between the approach of Refs.\
\cite{d,dk} and ours.

It should be remarked, however,
that in Ref.\ \cite{dk}
a conjecture was needed to use the translation
invariant measure for the Liouville mode
which is embedded as a component of the 2D
metric variables.
 In contrast to this, the Liouville mode $\phi$ in our approach
originates from the anomaly-compensating
degree of freedom, and
the functional measure in the path-integral quantization on the EPS
is fixed as the canonical measure
which is
translation invariant by construction.

In conclusion, our approach
to 2D gravity gives a common basis to formulate the theory in different
gauges, and is therefore suitable to study different
dynamical aspects of the theory.
\vspace{5mm}

\noindent
{\bf Acknowledgment}

We would like to thank K. Fujikawa, K. Itoh and H. Terao for useful
 discussions
and comments.

\end{document}